\documentclass[
    ,final            
  ]
  {aipproc}

\layoutstyle{6x9}

\begin{document}
\newcommand{\Od}{{\cal O}}
\newcommand{\lsim}   {\mathrel{\mathop{\kern 0pt \rlap
  {\raise.2ex\hbox{$<$}}}
  \lower.9ex\hbox{\kern-.190em $\sim$}}}
\newcommand{\gsim}   {\mathrel{\mathop{\kern 0pt \rlap
  {\raise.2ex\hbox{$>$}}}
  \lower.9ex\hbox{\kern-.190em $\sim$}}}
\title{Dark energy rest frame and the CMB dipole}


\author{Antonio L. Maroto}{
  address={Dept. F\'{\i}sica Te\'orica, Universidad Complutense de Madrid, 28040
Madrid, Spain}}

\begin{abstract}
 If dark energy can be described as a perfect fluid, then, apart from its 
equation of state
relating energy density and pressure, we should also especify 
the corresponding rest frame. Since dark energy is typically decoupled
from the rest of components of the universe, in principle such a frame could
be different from that of matter and radiation.   
In this work we consider the potential observable effects of the motion
of dark energy and the possibility to measure  the dark energy velocity
relative to matter. In particular we consider 
the modification of the usual interpretation
of the CMB dipole and its implications for the determination of matter
bulk flows on very large scales.  We also comment on the possible origin of a 
 dark energy flow and its evolution in different models. 
\end{abstract}

\maketitle


\section{Introduction}

 The content of the universe can be appropriately described as a four-fluid system:
baryons, radiation, dark matter and dark energy. Such fluids can be considered in 
good approximation to be decoupled from each other after recombination time. 
Each  fluid carries its own rest frame. Thus for instance, 
 the CMB radiation, being highly homogeneous allows us to define its rest frame 
 by means of the CMB dipole anisotropy. Indeed,
in the usual interpretation \cite{dipole}, the CMB dipole is due to the Doppler effect
caused by the motion of the observer with respect to the last scattering surface.
Therefore the radiation rest frame would be that in which an observer 
measures a vanishing dipole. 
This fact in turn has been used  to calculate the velocity of our Local Group 
with respect to
the CMB radiation, just by substracting the velocity of the Sun with respect to 
the Milky Way and the Milky Way velocity with respect to the Local Group.
The result is $v_{LG-CMB}=627\pm 22$ km s$^{-1}$ in the direction 
$(l,b)=(276\pm 3^\circ,30\pm 3^\circ)$ in galactic coordinates.

The rest frame of (dark) matter is not so easily defined since we know that matter
distribution is only homogeneous on very large scales. The presence of 
density inhomogeneities makes the velocity of a given matter volume to deviate
from the pure Hubble flow. Such deviations $v$ are known as peculiar velocities and
they are related to the density perturbations $\delta(x)$ by this simple expression to 
first order in perturbation theory:   
\begin{eqnarray}
\nabla \cdot v=-H_0\Omega_M^{0.6}\delta(x)
\end{eqnarray}
According to the Cosmological Principle, the universe is homogeneous on very
large scales and the amplitude of  density perturbations on scales of size $R$ will 
decline as we take larger and larger values of $R$. This in turn means that the 
peculiar velocity of a matter volume of radius $R$ will also decline
for larger $R$ (see Fig. 1), the 
matter rest frame eventually converging to the radiation frame on 
very large scales. In other words, matter and radiation would share a
common rest frame according to the Cosmological Principle. 

However as shown in Fig. 1, this theoretical framework is not
conclusively confirmed by  observations. Indeed, in recent years several
peculiar velocity surveys \cite{surveys} have tried to determine the volume size at which 
the streaming motion of matter with respect to the CMB vanishes. 
In  the figure the results
of different observations are compared with the {\it rms} 
expected bulk velocity
$V_b$ for  standard $\Lambda$CDM model in a sphere of radius $R$.
The results seem to agree with the theoretical expectations only at
scales $R\lsim 60h^{-1}$ Mpc. At larger scales, $R\gsim 100h^{-1}$ Mpc, 
different data sets lead to different bulk velocities both in amplitude
and direction. Moreover, there
are indeed measurements in which  large matter volumes are moving at 
speeds $\gsim 600$ km s$^{-1}$ with respect to the CMB frame, several
standard deviations away from the theoretical predictions. These results  
have been argued to be affected by systematic errors in distance
indicators, but if confirmed
by future surveys, a revision of some of the underlying ideas 
in  Standard
Cosmology would be required in order to understand the origin of
 such large flows.  

\begin{figure}
  \includegraphics[height=.4\textheight]{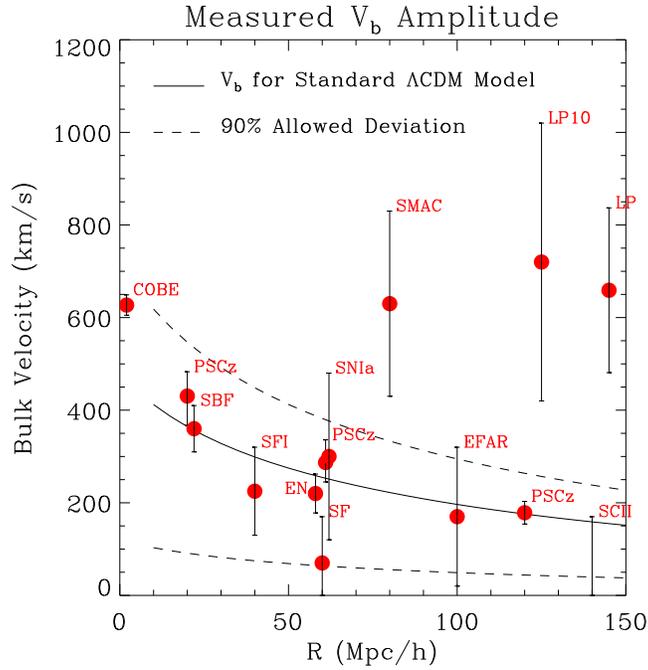}
\caption{\footnotesize Bulk  velocities vs. size $R$ for different peculiar
velocity surveys \cite{surveys} centered around the Local Group in the CMB frame.  
The dot labelled by 
COBE denotes the velocity of the Local Group measured by COBE. The solid line corresponds to the expected {\it rms}
velocity in the standard $\Lambda$CDM cosmology, together with the
90\% deviation in dashed lines. Figure from S. Zaroubi \cite{zaroubi}}
\end{figure}
   
Concerning the dark energy rest frame, it is usually assummed that dark energy
does not cluster on small scales and therefore can be considered as 
 a highly homogeneous fluid which is almost decoupled from the rest
of components, its only interaction being of gravitational nature. 
In such a case there is no reason to expect that its rest frame
should necessarily agree with the matter/radiation frame and therefore
an observational determination of the dark energy bulk velocity would be needed.  
In the following we will explore this problem, considering cosmological
models with four decoupled homogeneous fluids and different rest frames 
\cite{maroto}.

\section{Cosmology with moving dark energy}
Let us therefore consider  a cosmological scenario with four perfect fluids: 
baryons, radiation, dark matter and dark energy, whose equations of state read 
$p_\alpha=w_\alpha \rho_\alpha$ with  $\alpha=B,R,DM,DE$. For the sake
of generality, we will allow the dark energy equation of state to
have a smooth dependence on redshift $w_{DE}(z)$. 
The energy-momentum tensor of each fluid will take the form: 
\begin{eqnarray}
(T^\mu_{\;\;\nu})_\alpha=(\rho_\alpha+p_\alpha)
 u^\mu_\alpha u_{\nu\alpha}- p_\alpha\delta^\mu_{\;\;\nu}
\end{eqnarray}
Since  we are only interested in the effects of fluids
motion on the CMB dipole, it is sufficient to take into account
 the evolution of the largest-scale velocity
perturbations, i.e. we will just consider the zero-mode equations. The 
presence of inhomogeneities will contribute to higher multipoles. 
Therefore, for this particular problem we can write:
\begin{eqnarray}
\rho_\alpha&=&\rho_\alpha(\eta),\nonumber \\ 
 p_\alpha&=&p_\alpha(\eta), \nonumber \\ 
 u^\mu_\alpha&=&\frac{1}{a}(1,v^i_\alpha(\eta))
\end{eqnarray}    

We will assume that
$\vec v_\alpha^{\,2}\ll 1$ and we will work at first order 
in perturbation theory. In the particular case
we are considering, the form of the space-time 
metric will be given
by the following vector-perturbed spatially-flat 
Friedmann-Robertson-Walker metric:
\begin{eqnarray}
ds^2=a^2(\eta)\left(d\eta^2+2S_i(\eta)\,d\eta\, dx^i-\delta_{ij}
\,dx^i\,dx^j\right)
\label{metric}
\end{eqnarray}
 Accordingly,  
the total energy-momentum tensor reads:
\begin{eqnarray}
T^0_{\;\;0}&=&\sum_\alpha \rho_\alpha\nonumber \\
T^0_{\;\;i}&=&\sum_\alpha (\rho_\alpha+p_\alpha)(S_i-v_{i\alpha})\nonumber \\
T^i_{\;\;0}&=&\sum_\alpha (\rho_\alpha+p_\alpha)v^i_\alpha\nonumber \\
T^i_{\;\;j}&=&-\sum_\alpha p_\alpha\delta^i_{\;\;j}
\label{T}
\end{eqnarray}  
Notice that we are considering only the epoch after matter-radiation 
decoupling, assuming that dark energy is also decoupled and for that
reason we will ignore possible energy and momentum transfer
effects. 

We now calculate the linearized Einstein equations using
(\ref{metric}) and (\ref{T}).  They 
yield just the condition:
\begin{eqnarray}
S^i=\frac{\sum_\alpha (\rho_\alpha+p_\alpha)v^i_\alpha}
{\sum_\alpha (\rho_\alpha+p_\alpha)}
\label{S}
\end{eqnarray}

In General Relativity the combination 
$(\rho_\alpha+p_\alpha)$ 
appearing in (\ref{S}) 
plays the role of inertial mass density of the corresponding fluid, and 
accordingly $\vec S$ can be understood as the 
{\it cosmic center of mass velocity}. Notice that a pure cosmological
constant has no inertial mass density. 

 On the other hand, the energy conservation equations are trivially
satisfied, whereas from momentum conservation we see that the 
 velocity of each fluid relative to the center of mass frame scales
as: 
\begin{eqnarray}
\vert \vec S-\vec v_\alpha\vert\propto a^{3w_\alpha-1}
\label{velocity}
\end{eqnarray} 
Notice that for dark energy the scaling properties will depend 
on the particular
model under consideration \cite{maroto}.

Once we know the form of the perturbed metric, we can calculate the
effect of fluids motion on photons propagating from the last scattering
surface using standard tools \cite{Gio}. 
The energy of a photon coming from direction 
$n^\mu=(1,n^i)$ with $\vec n^{\,2}=1$ as seen by an observer moving with
velocity $u^\mu=a^{-1}(1,v^i)$ is given by $E=g_{\mu\nu}u^\mu P^\nu$, 
i.e. to first order in the perturbation:
\begin{eqnarray}
E\simeq \frac{\epsilon}{a}\left(1+\frac{d\delta x^0}{d\eta}+\vec n\cdot
(\vec S- \vec v)\right)
\end{eqnarray}
where $\epsilon$ parametrizes the photon energy and the perturbed 
trajectory of the photon reads 
$x^\mu(\eta)=x_0^\mu(\eta)+\delta x^\mu$, with $x_0^\mu=n^\mu \eta$. 

In order to obtain $d\delta x^0/d\eta$, we solve the geodesics equations
to first order in the perturbations, and for 
the 0-component we get
${d^2\delta x^0}/{d\eta^2}=0$. 
By defining $\hat{E}=a{E}$, the temperature fluctuation 
 reads:
\begin{eqnarray}
\left.\frac{\delta T}{T}\right\vert_{dipole}&=&
\frac{\hat{E}_0-\hat{E}_{dec}}{\hat{E}_{dec}}\simeq
\left.\frac{d\delta x^0}{d\eta}\right\vert^0_{dec}+\vec n\cdot
(\vec S- \vec v)\vert^0_{dec}\nonumber \\
&\simeq&\vec n\cdot
(\vec S- \vec v)\vert^0_{dec}
\label{dipolo}
\end{eqnarray}
where the indices $0$, $dec$ denote the present and decoupling times
respectively.

Today the only relevant contributions to the center of mass
motion are those of matter and dark energy, radiation  being negligible, so that:
\begin{eqnarray}
\vec S_0-\vec v_0\simeq\frac{\Omega_{M}(\vec v_{M}^0-\vec v_0)
+(1+w_{DE}^0)\Omega_{DE}
(\vec v_{DE}^0-\vec v_0)}
{1+w_{DE}^0\Omega_{DE}}
\end{eqnarray}  
where $w_{DE}^0$ is the present equation of 
state of dark energy and we have assumed that today the 
relative velocity of baryons and dark matter
is negligible. For that reason we have used $M$ to denote them both simultaneously. 
Notice that for a pure cosmological constant $w_{DE}^0=-1$
and there would be no contribution from dark energy in such a case.

At decoupling, the universe is matter dominated and we can
neglect the contribution to $\vec S$
from dark
energy.  Therefore:
\begin{eqnarray}
\vec S_{dec}-\vec v_{dec}
\simeq\frac{\Omega_{DM}}
{\Omega_{M}}(\vec v_{DM}^{dec}
-\vec v_B^{dec})\nonumber
\end{eqnarray}
where the emitter velocity is nothing but the baryonic 
velocity $\vec v_{dec}=\vec v_B^{dec}$. As we will see below, 
when dark energy is absent, it is in the form of a pure cosmological constant  
or it is at rest with respect to 
radiation,  
we will have $\vec v_{DM}^{dec}=\vec v_B^{dec}$
 and this term vanishes. 

According to this result, the CMB dipole has two different types of 
contributions: one
is the usual Doppler effect due to the change of velocity between 
emitter and observer; the second contribution comes from the fact that the
photon is propagating in an anisotropic medium which is changing in time. This
second contribution is precisely given by the change in the velocity of the
cosmic center of mass between emission and reception. 
The possibility that the dipole is not enterely due to a Doppler effect has been
considered previously in the litereature in different contexts (see \cite{nokin}).

 When all the components share a common rest frame 
then the previous result reduces to the usual expression for the dipole:
$\delta T/T\vert_{dipole}\simeq 
\vec n\cdot(\vec v_R^0- \vec v_0)$. 
However in general
it is possible that an observer at rest with  radiation 
$\vec v_0=\vec v_R^0\neq\vec v_M^0 \neq \vec v_{DE}^0$ can measure an
nonvanishing dipole according to (\ref{dipolo}).

In the absence of dark energy or in the case in which it is in the 
form of a pure 
cosmological constant ($w_{DE}=-1$), dark energy would not contribute 
to the center of mass motion. Moreover, today the radiation contribution
is negligible and accordingly the center of mass
rest frame would coincide with the matter rest frame.
This implies that  the relative motion of matter and radiation
today could not 
explain the existence of  bulk flows on the largest scales, since the 
frame in which the dipole vanishes would coincide with the matter
rest frame.  Conversely, the 
existence of non-vanishing bulk flows would require the presence 
of moving dark energy with $w_{DE}^0\neq -1$. 

Indeed, if moving dark energy is responsible for the 
existence of cosmic bulk flows  on very large scales, 
then the amplitude and direction of such flows would 
provide a way to measure the relative velocity of matter and dark energy. 
As commented
above, the bulk flow $\vec V_b$ 
can be
understood as the average velocity of a given matter volume with respect
to an observer  who  measures a vanishing CMB dipole, i.e. 
$\vec V_b=\vec v_M^0-\vec v_0$. Such an 
observer has a  velocity which can be obtained from (\ref{dipolo}), 
and accordingly:
\begin{eqnarray}
\vec V_b \simeq \frac{(1+w_{DE}^0)\Omega_{DE}}
{1+w_{DE}^0\Omega_{DE}}
(\vec v_M^0-\vec v_{DE}^0)
+\frac{\Omega_{DM}}{\Omega_M}(\vec v_{DM}^{dec}-\vec v_B^{dec}) \nonumber
\end{eqnarray}

Notice that, according to these results, 
even if matter is at rest with respect to the CMB radiation, 
$\vec v_M^0=\vec v_R^0$, 
it would be possible to have a non-vanishing flow $\vec V_b\neq 0$, provided
dark energy is moving with respect to matter. 

\section{A primordial dark energy flow?}
So far we have considered only the effects of the different components of
the universe having different rest frames. However we have not discussed 
what is the origin of such a velocity offset.
In standard cosmology, baryonic matter and radiation were coupled before 
recombination, and this was also so for dark matter, 
in the case in which it is in the form of weakly interacting massive 
particles. However the nature of dark energy is still a mistery and 
we ignore what is the type of interaction, if any, between dark energy and 
radiation/matter. For that reason, the primordial value of the dark energy rest 
frame velocity with 
respect to radiation/matter should be  considered as a  
completely free cosmological parameter, which could be determined only 
by observations.  

\begin{figure}
  \includegraphics[height=.4\textheight]{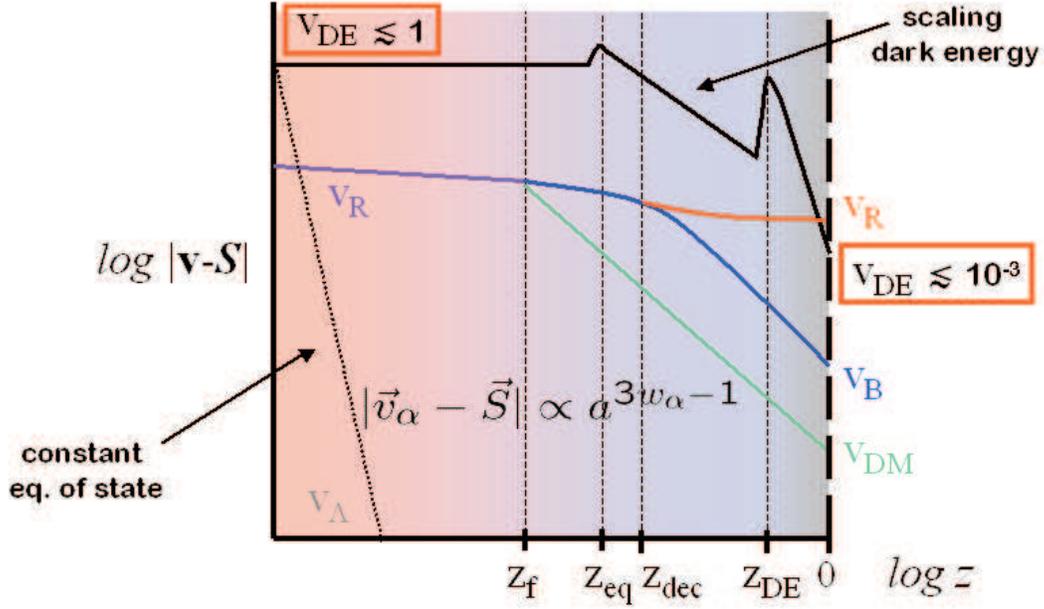}
\caption{\footnotesize Qualitative evolution of  fluids velocities in 
the cosmic center of mass
frame for different dark energy models. The black full line corresponds to scaling
dark energy. The various jumps arise because of 
momentum conservation in the corresponding transitions between different eras. 
  The dotted black line is the $w_{DE}\simeq -1$ model.}
\end{figure}

Although the 
primordial value is unknown, its subsequent evolution is given by 
(\ref{velocity}).
In Fig. 2 we show the evolution of the velocities of the different fluids 
with respect to the center of mass frame. 
We compare two models for dark energy, one in which the  equation  
of state is a constant close to $-1$ and a scaling \cite{scaling}
dark energy model
in which the equation of state mimics that of the 
dominant component. We see that the velocity of radiation is almost constant
throughout the cosmological evolution,   
whereas although  matter  is initially dragged by radiation,  its velocity starts
decaying as $a^{-1}$  after decoupling. Dark energy velocity is damped 
very fast for constant equation of state, whereas it could have appreciable 
amplitude today in scaling models.    

Apart from the effects on the dipole, the metric anisotropies generated by the 
motion of dark energy can give rise in some cases to non-negligible quadrupole
contributions and generate a net polarization of the CMB radiation. Such effects
 offer additional possibilities for the determination of  
the dark energy rest frame. These
results will be presented elsewhere \cite{BM}.

\vspace{.5cm}

{\em Acknowledgments:} 
This work has been partially supported by DGICYT (Spain) under project numbers
FPA 2004-02602 and FPA 2005-02327

\end{document}